\documentclass{aa}

\usepackage{graphicx}
\usepackage{txfonts}
\usepackage{textcomp}
\usepackage{epstopdf}
\usepackage{subfigure}
\usepackage{natbib}
\usepackage[utf8]{inputenc}
\usepackage{german}
\usepackage{color}

\usepackage{marginnote}

\begin{document}

   \title{What drives the dust activity of comet 67P/Churyumov-Gerasimenko?}

   \subtitle{}

   \author{B. Gundlach\inst{1}, J. Blum\inst{1}, H. U. Keller\inst{1}, Y. V. Skorov \inst{2}}

   \institute{Technische Universit{\"a}t Braunschweig, Institut f{\"u}r Geophysik und extraterrestrische Physik, Mendelssohnstra{\ss}e 3, D-38106 Braunschweig, Germany\\
              \email{b.gundlach@tu-bs.de}
         \and
             Max-Planck-Institut f{\"u}r Sonnensystemforschung, Justus-von-Liebig-Weg 3, D-37077 G{\"o}ttingen, Germany\\
             }

   \date{Received , Accepted }

% \abstract{}{}{}{}{}
% 5 {} token are mandatory

\abstract{The gas-driven dust activity of comets is still an unsolved question in cometary physics. Homogeneous dust layers composed of micrometer-sized grains possess tensile strengths of $\sim 1 \, \mathrm{kPa}$, which is far higher than typical gas pressures caused by the sublimation of the ices beneath the covering dust layer. This implies that the dust grains cannot be detached from the surface by the gas pressure of the sublimating ices. One possibility to avoid this problem is that the nucleus formed through the gravitational collapse of an ensemble of millimeter- to centimeter-sized aggregates. In this case, an aggregate layer with a tensile strength on the order of $\sim 1 \, \mathrm{Pa}$ is formed on the surface of the nucleus, which allows for the release of the aggregates from the surface by the gas pressure build up at the ice-dust interface.}
{We use the gravitational instability formation scenario of cometesimals to derive the aggregate size that can be released by the gas pressure from the nucleus of comet 67P/Churyumov-Gerasimenko for different heliocentric distances and different volatile ices.}
{To derive the ejected aggregate sizes, we developed a model based on the assumption that the entire heat absorbed by the surface is consumed by the sublimation process of one volatile species. The calculations were performed for the three most prominent volatile materials in comets, namely, $\mathrm{H_20}$ ice, $\mathrm{CO_2}$ ice, and $\mathrm{CO}$ ice.}
{We find that the size range of the dust aggregates able to escape from the nucleus into space widens when the comet approaches the Sun and narrows with increasing heliocentric distance, because the tensile strength of the aggregates decreases with increasing aggregate size. The activity of $\mathrm{CO}$ ice in comparison to $\mathrm{H_20}$ ice is capable to detach aggregates smaller
by approximately one order of magnitude from the surface. As
a result of the higher sublimation rate of $\mathrm{CO}$ ice, larger aggregates are additionally able to escape from the gravity field of the nucleus.}
{Our model can explain the large grains (ranging from $2\,\mathrm{cm}$ to $1\,\mathrm{m}$ in radius) in the inner coma of comet 67P/Churyumov-Gerasimenko
that have been observed by the OSIRIS camera at heliocentric distances between $3.4\, \mathrm{AU}$ and $3.7\, \mathrm{AU}$. Furthermore, the model predicts the release of decimeter-sized aggregates (trail particles) close to the heliocentric distance at which the gas-driven dust activity vanishes. However, the gas-driven dust activity cannot explain the presence of particles smaller than $\sim 1 \mathrm{mm}$ in the coma because the high tensile strength required to detach these particles from the surface cannot be provided by evaporation of volatile ices. These smaller particles can be produced for instance by spin-up and centrifugal mass loss of ejected larger aggregates.}

%  \abstract{}
%  {}
%  {}
%  {We find that the ; bis mm erklrbar; große agg kein problemunterschied $\mathrm{H_2O}$, $\mathrm{CO_2}$ and $\mathrm{CO}$}
%  {}

   {}
   {}
   {}
   {}
   {}


   \maketitle
%
%________________________________________________________________
\section{Introduction}\label{Introduction}
Comets and their activity are most often associated with the emission of dust that forms a visible coma and a tail, the shape of the latter determined by the radiation pressure of the sun. The role of volatiles for cometary activity, first observed by the emission of radicals but not of stable molecules, was initially strongly underestimated, which led to models such as the \textit{\textup{sandbank}} model \citep{Lyttleton1953}. Better spectroscopic observations and the determination of nongravitational forces led to the \textit{\textup{dirty snowball}} model \citep{Whipple1950,Whipple1951}, where ices such as solid $\mathrm{H_2O}$, $\mathrm{CO_2}$ , and $\mathrm{CO}$, polluted by dust, form a solid nucleus. The dust is then liberated by the sublimation of the abundant volatile ices. The first space missions to comet 1P/Halley \citep{Sagdeev1986,Keller1986} revealed a pitch-black nucleus with a density of only $0.6^{+0.9}_{-0.4} \, \mathrm{g\, cm^{-3}}$. It was determined that dust, that is, a refractory material, \textbf{is more abundant than} volatiles. In situ measurements and estimates revealed a dust-to-gas mass ratio somewhere between 1 and 10. The observations of comet 1P/Halley also showed that only a small part of its surface was active. Most of the surface seemed to be covered by an inert crust for which cohesive forces between the dust grains are much stronger than gravitational forces \citep{Heim1999}. The question how activity works and how it is maintained over many cometary orbits  remained unanswered.
\par
Modeling efforts concentrated on the heat transport into the porous nucleus, which is required to sublime the ices. The material structure (porosity and packing structure) and the physical properties (e.g., heat conductivity) strongly influence the gas production rates \citep{Gundlach2011}. Modeling parameters were adjusted to fit the observations \citep[see, e.g.,][]{Prialnik2004, Enzian1999}. In the models, the dust was somehow liberated by the gas, leaving the surface, and was accelerated in the inner coma against the minute gravity of the nucleus \citep{FinsonProbstein1968}. \citet{KuehrtKeller1994} showed that the sublimation vapor pressure is much too weak to overcome the van-der-Waals forces between small dust grains. Following this paper, \citet{Moehlmann1995} presented typical values of cohesion of porous media and concluded that a cohesively bound matrix of relatively refractory and porous matter cannot be destroyed by the gas drag. To work around this insuperable obstacle, the cohesion force between microscopic mineral grains was just ignored in most similar publications \citep[see, e.g.,][]{Rickman1990}. Once this dust crust has formed, activity is quenched by the reduction of both the heat flow into the interior and the gas flux from the volatile subsurface region of the cometary nucleus.
\par
Recently, however, it was shown that comets can form by gravitational collapse of centimeter-sized dust agglomerates in the early solar system \citep{Johansen2007,Skorov2012,Blum2014,Blum2015}. There are strong indications that the dust and water-ice agglomerates were mixed by radial transport of the particles before the cometesimals formed in the outer solar system beyond the CO sublimation boundary \citep[see results of the Stardust mission;][]{Brownlee2006}. The gravitational collapse then forms a cometesimal composed of porous dust and ice aggregates, themselves consisting of micrometer-sized particles \citep{Sunshine2007,Sunshine2011,Protopapa2014}. The existence of separate ice and dust aggregates composed of micron-sized particles is also suggested by the observations of grains containing water ice in the inner coma of comet 103P/Hartley 2 \citep{AHearnetal2011} and on the surface of comet 9P/Tempel 1 \citep{Sunshine2006}.
\par
When the comet approaches the Sun for the first time, the ice on the surface sublimes and the remaining dust agglomerates form a thin, non-volatile dust layer \citep{Skorov2012}. The ice-free surface material possesses an ultra-low tensile strength (i.e., the cohesive force per unit area) of only a few Pascal, or less in the case of millimeter-sized aggregates \citep{Blum2014}. This is similar to the highest gas pressures that can be reached by the outgassing of typical volatile constituents such as $\mathrm{H_2O}$ ice, $\mathrm{CO_2}$ ice, or $\mathrm{CO}$ ice, at typical subsurface temperatures of comet nuclei \citep{Blum2014,Blum2015}. The thin layer of dust aggregates can then be ejected if the gas pressure beneath it exceeds the tensile strength of the material. The vapor pressure of $\mathrm{H_2O}$ ice suffices to support the dust activity of such a gently formed nucleus at heliocentric distances smaller than $2.5 \, \mathrm{AU}$ \citep{Blum2015}. The aggregates are released intact because of their relatively high internal tensile strength \citep{Blum2006a,Kothe2010}. Icy aggregates will have to be lifted with the help of super-volatiles (e.g., $\mathrm{CO_2}$ ice, or $\mathrm{CO}$ ice) in agreement with the above-mentioned observations in the inner coma of comet Hartley 2. \citet{Blum2014} have already shown that the ejection of $\mathrm{H_2O}$ ice aggregates is formally possible by the outgassing of $\mathrm{H_2O}$ ice at $1 \, \mathrm{AU}$ (see their Fig. 10). However, as this is physically not possible, materials possessing higher volatility are required to release $\mathrm{H_2O}$ ice aggregates.
\par
At larger heliocentric distances, the sublimation pressure of $\mathrm{H_2O}$ ice is too low to explain the dust emission of comets, however. Beyond $2.5 \, \mathrm{AU}$, more volatile icy compounds are needed to explain the observed activity of comets. $\mathrm{CO_2}$ and in particular $\mathrm{CO}$ ice possesses a lower latent heat (i.e., a higher sublimation rate if the same amount of energy is available for sublimation) than $\mathrm{H_2O}$. Indeed, $\mathrm{CO_2}$ and $\mathrm{CO}$ are frequently observed in relatively high abundances of $\sim10\, \%$ to $\sim20\, \%$ with respect to $\mathrm{H_2O}$ \citep[see, e.g.,][]{BockeleeMorvan2004,AHearn2012}. At these heliocentric distances, $\mathrm{H_2O}$ ice is inactive, so that the outgassing of $\mathrm{CO_2}$ ice, or $\mathrm{CO}$ ice causes the ejection of $\mathrm{H_2O}$ ice aggregates. The ejection of $\mathrm{H_2O}$ ice aggregates by the activity of $\mathrm{CO_2}$, or $\mathrm{CO}$ ice is also possible at shorter heliocentric distances, however, as can be seen in the case of comet 103P/Hartley 2 \citep{AHearnetal2011}.
\par
The observed $\mathrm{CO}$ production rate of comet Hale-Bopp (C/1995 O1) directly followed the variation of the solar irradiation from $5 \, \mathrm{AU}$ down to perihelion and out again \citep{Biver1999,Gortsas2011}. This suggests that $\mathrm{CO}$ ice must be present relatively close to the surface of the nucleus even at perihelion. In the case of comet 103P/Hartley 2, the outgassing of $\mathrm{CO_2}$ ice was observed to drive the ejection of centimeter-sized icy particles \citep{AHearnetal2011}. These observations strongly support the hypothesis that super-volatiles can survive close to the eroding surface and might be real drivers of observable dust activity.
\par
To fit numerous photometric and colorimetric observations of comet dust \citep{Kolokolova2007}, cometary grains have to be porous aggregates. Comet polarimetry in particular has shown that these aggregates in turn consist of micron and submicron porous pieces. This structure of the cometary particles leads to a dramatic reduction of the van-der-Waals forces between the contacting aggregates \citep{Skorov2012}. As noted previously \citep{Skorov2012,Blum2014}, it is easier to lift dust layers consisting of large agglomerates than those fromed by small ones. Thus the evaluation of cohesion between the aggregates provides the lower limit of the size of the ejected aggregates. On the other hand, the gravity of the comet restricts the maximum size of the aggregates that can be accelerated away by the out-flowing gas \citep{Whipple1950}. The early observations in the inner coma of comet 67P/Churyumov-Gerasimenko by the Rosetta spacecraft revealed a considerable amount of large particles. Images taken by the OSIRIS camera \citep{Kelleretal2007} even at heliocentric distances larger than $3.6\, \mathrm{AU}$ are full of tracks of particles of up to $2\, \mathrm{cm}$ in diameter \citep{Rotundi2015}. Furthermore, particles with diameters between $4\,\mathrm{cm}$ and $2\,\mathrm{m}$ were observed in bound orbits close to the nucleus of comet 67P/Churyumov-Gerasimenko between $3.4\, \mathrm{AU}$ and $3.7\, \mathrm{AU}$. These bound particles were possibly emitted during or after the last perihelion passage of comet 67P/Churyumov-Gerasimenko.
\par
In this work, we discuss how the gas-driven dust activity of comets can be modeled based on laboratory results and on the assumption that comets have formed by gravitational instability (see Sect. \ref{Modeling}). In Sect. \ref{Application}, this model is then applied to comet 67P/Churyumov-Gerasimenko to show that the outgassing of $\mathrm{CO_2}$ ice and $\mathrm{CO}$ ice preferentially ejects large aggregates, which has been observed by Rosetta, and that evaporating $\mathrm{H_2O}$ ice cannot eject any particles at large heliocentric distances ($>2.5\,\mathrm{AU}$). Section \ref{Summary} summarizes the main results of this work.

\section{Modeling cometary dust activity}\label{Modeling}
    \begin{figure*}[p!]
    \centering
    \includegraphics[angle=0,width=1\textwidth]{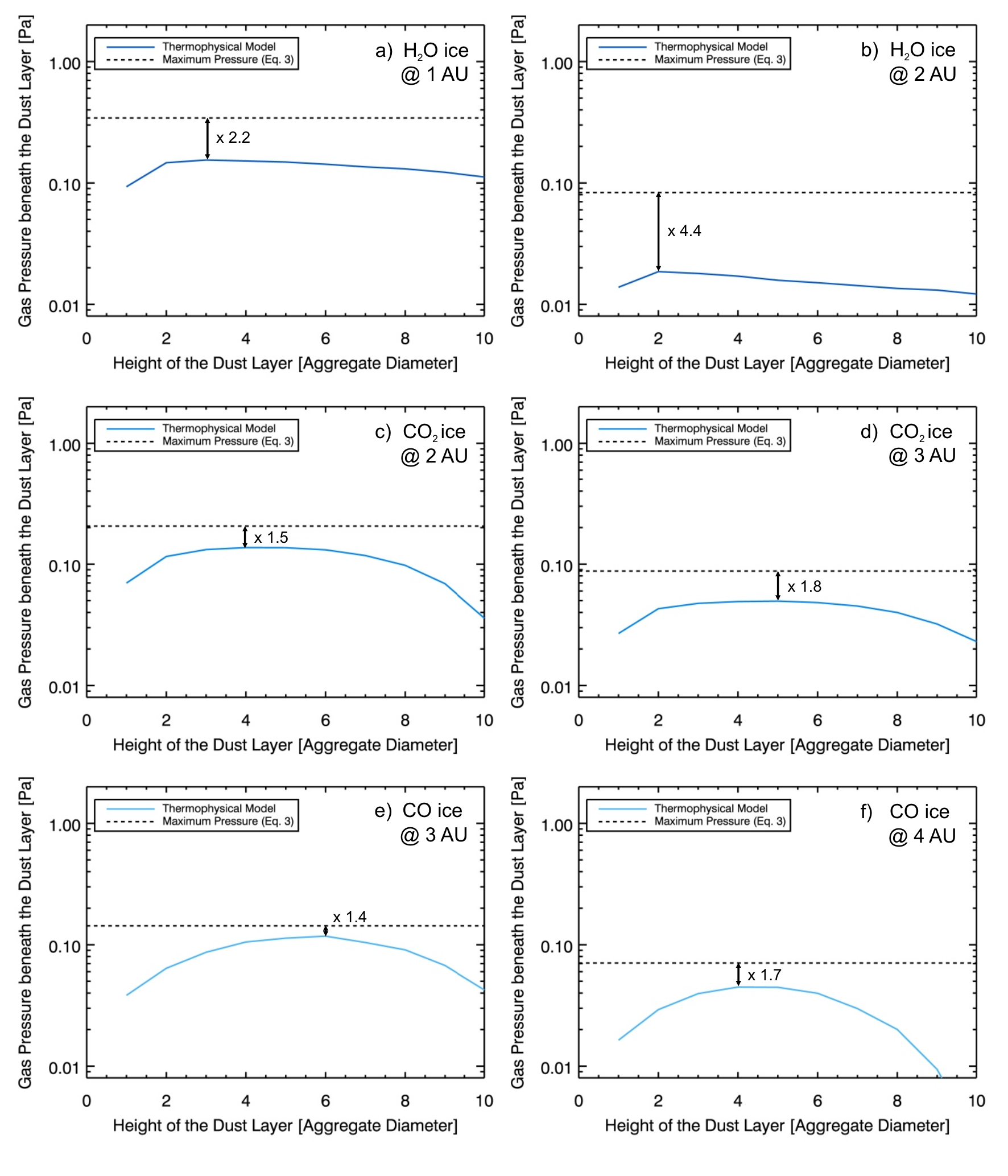}
    \caption{Gas pressure beneath the covering dust layer as a function of the thickness of the dust layer (in units of the dust-aggregate diameter) at different heliocentric distances (between $1 \, \mathrm{AU}$ and $4 \, \mathrm{AU}$) and for different volatile materials ($\mathrm{H_2O}$ ice, $\mathrm{CO_2}$ ice and $\mathrm{CO}$ ice). The solid blue curves and the black dashed lines show the gas pressure derived by the thermophysical model and the approximation by Eq. \ref{eq_pressure}, respectively. The temperatures of the ice surface at the peak of the pressure curves are $208.4 \, \mathrm{K}$ ($\mathrm{H_2O}$ ice at $1 \, \mathrm{AU}$), $196.3 \, \mathrm{K}$ ($\mathrm{H_2O}$ ice at $2 \, \mathrm{AU}$), $117.2 \, \mathrm{K}$ ($\mathrm{CO_2}$ ice at $2 \, \mathrm{AU}$), $106.6 \, \mathrm{K}$ ($\mathrm{CO_2}$ ice at $3 \, \mathrm{AU}$), $33.9 \, \mathrm{K}$ ($\mathrm{CO}$ ice at $3 \, \mathrm{AU}$), and $32.8\, \mathrm{K}$ ($\mathrm{CO}$ ice at $4 \, \mathrm{AU}$). The deviation of the approximation from the highest pressures derived by the thermophysical model are denoted by the arrows. The sublimation pressure is calculated by $\mathrm{ln}(p_{sub}) = a \,-\, b / T$, with the coefficients $a=3.23\times10^{12} \, \mathrm{Pa}$ and $b=6134.6\,\mathrm{K}$ taken from \citet{Gundlach2011} for $\mathrm{H_2O}$ ice. For $\mathrm{CO_2}$ ice and $\mathrm{CO}$ ice the coefficients are $a=1.23\times10^{12} \, \mathrm{Pa}$ and $b=3167.8\,\mathrm{K}$ as well as $a=1.26\times10^{9} \, \mathrm{Pa}$ and $b=764.2\,\mathrm{K}$, respectively \citep{FanaleSalvail1990}. We used a Bond albedo of $0.04$ for the calculations \citep[the geometric albedo of comet 67P/Churyumov-Gerasimenko is $0.05$;][]{Sierks2015}.}
    \label{fig_pressures}
    \end{figure*}
To release dust particles from a cometary nucleus, the force exerted on the dust by the outflowing gas molecules must exceed the cohesive force of the dust aggregates. Here, we recall that the assumption of an ice-free surface is an idealization of the problem, but observations have shown that most of the cometary surfaces are covered by a nonvolatile dust layer \citep[see, e.g.,][]{Keller1988, Buratti2004}. The tensile strength of the dust-aggregate layer reads
\begin{equation}
Y \, = \, Y_0 \, \phi \, s^{-2/3} \, \mathrm{,}
\label{eq_tensile}
\end{equation}
where $Y_0 \, = \, 1.6 \, \mathrm{Pa}$ is a constant, $\phi$ is the volume filling factor of the aggregate layer, and $s$ is the dust-aggregate radius measured in millimeters \citep{Skorov2012}. \citet{Brissetetal2015} experimentally showed that Eq. \ref{eq_tensile} is quantitatively correct (see their Fig. 17). The smallest dust-aggregate radius that can be released from the surface by the sublimation of the ices is given if the gas pressure $p$  equals the tensile strength of the aggregate layer (Eq. \ref{eq_tensile}),
\begin{equation}
s_{min} \, = \, \left(\frac{p}{Y} \, \frac{1}{\phi}\right)^{-3/2}  \, \mathrm{.}
\label{eq_lower}
\end{equation}
For cometary surface layers formed by gravitational instability (i.e., for dust aggregate layers), the tensile strength is very
low ($\sim 1 \, \mathrm{Pa}$) and somewhat counterintuitively decreases with increasing dust-aggregate size \citep{Skorov2012,Blum2014}. Hence, it is easier to release larger dust aggregates than smaller ones. For example, a densely packed aggregate layer with a volume-filling factor of $0.6$ consisting of $100 \, \mathrm{\mu m}$-sized dust aggregates possesses a tensile strength of $4.5 \, \mathrm{Pa}$, whereas increasing the aggregate size to $1 \, \mathrm{cm}$ yields a tensile strength of $0.2 \, \mathrm{Pa}$.
\par
The gas pressure $p$ of the sublimating ice, covered by a porous dust layer, depends on its transport characteristics (gas permeability) and on the energy supplied to the region. The latter depends on the heliocentric distance, on the nucleus albedo, on the thickness, and on the heat conductivity of the covering dust layer. A detailed heat-conductivity model for granular materials in vacuum was presented by \citet{Gundlach2012} and \citet{GundlachBlum2013}. Together with the available energy for the sublimation process, the permeability of the covering dust layer determines the pressure build-up at the depth of the ice-dust interface \citep[][]{Gundlach2011,Blum2014,Blum2015}. In the following, this detailed model is referred to as the thermophysical model.
\par
In general, the derivation of the gas pressure at the ice-dust interface should be determined from full consideration of the conservation of energy, momentum, and mass for solid and gaseous phases. However, such a detailed consideration is far beyond the scope of this paper. We wish to simplify the physical model as much as possible (keep it reasonable) to highlight the new qualitative results. For this reason, we simplified the derivation of the gas pressure at the ice-dust interface by the approximation that the incoming energy that is absorbed by the nucleus is entirely used for the sublimation of the volatiles (i.e., thermal reradiation of the comet nucleus and energy dissipation by its core are neglected). The reasonableness of this simplification is illustrated below. With this assumption, the gas pressure at the ice-dust interface is given by
\begin{equation}
p \, = \, \left( \, 1 \, - \, A \, \right) \, I_{E} \, \left( \, \frac{a_E}{a_h} \, \right)^2 \, \frac{1}{\Lambda} \, \sqrt{\frac{2 \, \pi \, k \, T_{ice}}{m}} \, \mathrm{,}
\label{eq_pressure}
\end{equation}
where $A$ is the Bond albedo of the surface material, $I_{E}$ is the solar constant, $a_E$ is the distance of the Earth to the Sun, $a_h$ is the heliocentric distance of the comet, $\Lambda$ is the latent heat of the sublimating ice, $k$ is the Boltzmann constant, $T_{ice}$ is the temperature of the evaporating ice, and $m$ is the mass of the evaporating molecules.
\par
To illustrate the plausibility of this simplification, the results of Eq. \ref{eq_pressure} are compared with the corresponding pressure obtained from the sophisticated thermophysical model for $\mathrm{H_2O}$ ice \citep[latent heat of sublimation: $\Lambda = 2.6\times 10^6 \, \mathrm{J \, kg^{-1}}$;][]{Orosei1995} at the ice-dust interface for different thicknesses of the covering dust-aggregate layer and at two different heliocentric distances (at $1 \, \mathrm{AU}$ and $2 \, \mathrm{AU}$; see Figs. \ref{fig_pressures}a and \ref{fig_pressures}b).
\par
For another test of the validity of the approximation, we calculated the resulting gas pressure for $\mathrm{CO_2}$ ice \citep[at $2 \, \mathrm{AU}$ and $3 \, \mathrm{AU}$; latent heat of sublimation: $\Lambda=5.7\times 10^5 \, \mathrm{J \, kg^{-1}}$;][see Figs. \ref{fig_pressures}c and \ref{fig_pressures}d]{Mavko2009} and for $\mathrm{CO}$ ice \citep[at $3 \, \mathrm{AU}$ and $4 \, \mathrm{AU}$; latent heat of sublimation: $\Lambda=3.0\times 10^5 \, \mathrm{J \, kg^{-1}}$;][see Figs. \ref{fig_pressures}e and \ref{fig_pressures}f]{BrownZiegler1979}. These two volatile materials were chosen because of their relatively high abundance in comets and because model calculations suggest that volatile species incorporated in planets after their formation are mainly composed of $\mathrm{H_2O}$, $\mathrm{CO_2}$, $\mathrm{CO}$, $\mathrm{CH_3OH,}$ and $\mathrm{NH_3}$ \citep{Marboeuf2014}.
\par
A comparison of the derived gas pressure at the ice-dust interface using the thermophysical model (solid blue curves in Fig. \ref{fig_pressures}) with the approximation of Eq. \ref{eq_pressure} (dashed black line; derived for the temperature of the ice-dust interface at which the highest pressure is reached in the thermophysical model) yields the deviations indicated by the arrows in Fig. \ref{fig_pressures}. The pressure ratio between the approximation and the thermophysical model indicated in Fig. \ref{fig_pressures} was calculated for the maxima of the pressure curves, because the maximum yields the smallest aggregate size that can be released by the gas pressure. For the considered cases, the maximum gas pressure is typically reached for a dust layer thickness between two and five aggregate layers.
\par
Figure shows \ref{fig_pressures} that the deviation from the thermophysical model to the approximation of Eq. \ref{eq_pressure} increases with increasing heliocentric distance. This can be explained by the following considerations. For reasonable dust-aggregate sizes and surface temperatures, the heat conductivity is dominated by radiative transport and, thus, increases strongly with increasing temperature, that is, with decreasing heliocentric distance \citep[see][for details]{Gundlach2012,GundlachBlum2013}. Therefore, the fraction of absorbed energy that is transported into the interior of the comet nucleus and, thus, used to evaporate volatile ices, also increases with decreasing heliocentric distance. On the other hand, for increasing heliocentric distances, the relative importance of thermal reradiation of the surface into deep space increases \citep[see][for details]{Blum2015}.
\par
For example, the smallest aggregate size increases from $2 \, \mathrm{mm}$ to $6 \, \mathrm{mm}$ for $\mathrm{H_2O}$ ice activity at $1 \, \mathrm{AU}$ if the thermophysical model is used instead of the approximation. At $2 \, \mathrm{AU}$ and for $\mathrm{H_2O}$ ice sublimation, the smallest aggregate size changes from $18 \, \mathrm{mm}$ (approximation) to $174 \, \mathrm{mm}$ (thermophysical model).
\par
\begin{figure}[t!]
        \centering
        \includegraphics[angle=180,width=1\columnwidth]{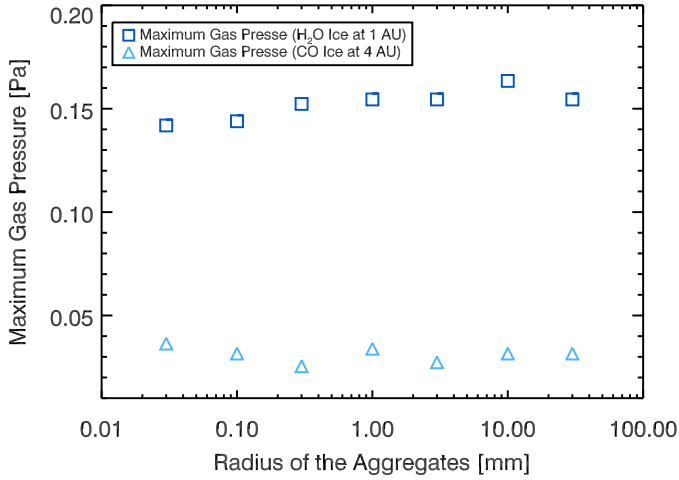}
        \caption{Highest gas pressure as a function of the radius of the dust aggregates covering the dust-ice interface. The calculations are performed for $\mathrm{H_2O}$ ice at $1 \, \mathrm{AU}$ (blue squares) and for $\mathrm{CO}$ ice at $4 \, \mathrm{AU}$ (blue triangles). These two examples are chosen to cover the closest and the farthest heliocentric distance investigated with the thermophysical model (see Fig. \ref{fig_pressures}).}
        \label{fig_pmax_s}
        \end{figure}

The calculations shown in Fig. \ref{fig_pressures} were only
performed for one aggregate size of $1 \, \mathrm{mm}$  because the systematic simulations showed that starting from an aggregate radius of $15 \, \mathrm{\mu m,}$ the variation of the aggregate size does not considerably influence the resulting pressure at the ice-dust interface (see Fig. \ref{fig_pmax_s}).
\par
With the knowledge of the gas pressure at the ice-dust interface, it is possible to derive the smallest dust-aggregate size that can be released from the surface by comparing the gas pressure with the tensile strength of the dust-aggregate layer.
\par
Strictly speaking, the assumption that all absorbed energy is used by the sublimation of the ice gives us the highest effective sublimation rate (i.e., highest mass loss per second) and not pressure. However, numerical tests agree with the values retrieved from Eq. \ref{eq_pressure} (see Fig. \ref{fig_pressures}). Considering the thermal emission in the energy budget leads to a reduction of the effective loss mass. The role of this dissipation channel was discussed by \citet{Blum2015}, where it was shown that this effect is weak close to the Sun and increases with heliocentric distance. Because the tensile strength of the covering aggregate layer is a decreasing function of aggregate size (see Eq. \ref{eq_tensile}), the larger grains can be lifted easily and their size is only restricted by the cometary gravity. In addition, the gas drag is stronger for larger aggregates because their cross section is larger. Thus, the application of Eq. \ref{eq_lower} together with Eq. \ref{eq_pressure} yields the smallest dust-aggregate size that can be released by the gas pressure at the ice-dust interface. For the presented cases shown in Fig. \ref{fig_pressures}, the thermophysical model provides lower values for gas pressure and, hence, higher values for the smallest dust-aggregate size, because it takes into account other channels of the incoming energy dissipation. Nonetheless, for $\mathrm{H_20}$ ice, $\mathrm{CO_2}$ ice, and $\mathrm{CO}$ ice, the approximation of the highest gas pressure (Eq. \ref{eq_pressure}) seems reasonable (see Fig. \ref{fig_pressures}).
\par
The dust aggregates are lifted up by the outward-directed gas-friction force, which can be approximately described by $F_{gas} \, = \, \pi \, s^2 \, p \,  (R/r)^2$ \citep{Blum2015}. Obviously, to release aggregates from the cometary surface, the gas drag force has to exceed the inward-directed gravitational force, $F_{grav} \, = \, G \, (m \, M \, / \, R^2) \, (R/r)^2$, that
is, $F_{gas} \geq F_{grav}$. Here, $R$ is the radius of the comet nucleus (for simplicity, we assumed a spherical shape for the nucleus), $r$ is the distance between the dust aggregate and the center of the nucleus, $G$ is the gravitational constant, $m$ is the mass of the dust aggregate, and $M$ is the mass of the nucleus. Neglecting cohesion, we calculated the largest aggregate that can be ejected from the nucleus,
\begin{equation}
s_{max} \, = \, \frac{9}{16 \, \pi \, G} \, \frac{p}{\rho_{comet} \, \rho_{aggregate} \, R } \, \mathrm{.}
\label{eq_upper}
\end{equation}
Here, $\rho_{comet}$ and $\rho_{aggregate}$ are the mass densities of the comet and the dust aggregates. We used an aggregate density of $875 \, \mathrm{kg\, m^{-3}}$ (volume-filling factor of the aggregates multiplied by the bulk density). Equation \ref{eq_upper} is only valid under the crude assumption that the nucleus and the released dust aggregate are spherical in shape and that the gas release is constant over the surface of the comet nucleus. Dust aggregates larger than the estimated size in Eq. \ref{eq_upper} cannot be ejected into space.

\section{Application of the model to comet 67P/Churyumov-Gerasimenko}\label{Application}
In this section, we apply the model described in Sect. \ref{Modeling} to comet 67P/Churyumov-Gerasimenko to derive the size range of the dust aggregates able to escape from the nucleus by the outgassing of icy materials. The lower limit of the dust-aggregate size is determined by the tensile strength of the material that has to be exceeded by the gas pressure at the ice-dust interface (Eqs. \ref{eq_lower} and \ref{eq_pressure}). The upper limit is given by the fact that the released aggregates have to escape the gravity field of the nucleus by the gas drag (outward-directed gas-friction force; see Eq. \ref{eq_upper}).
\par
The calculations were performed for $\mathrm{H_2O}$ ice, $\mathrm{CO_2}$ ice and $\mathrm{CO}$ ice. Each icy material was treated by a one-species model, which means that all available energy was consumed by the sublimation of this particular icy species. To derive the largest dust-aggregate size, a radius of $2 \, \mathrm{km}$ \citep{Lowry2012} and a density of $470 \, \mathrm{kg\, m^{-3}}$ \citep{Sierks2015} were used for the nucleus of comet 67P/Churyumov-Gerasimenko. Furthermore, we used a Bond albedo of $0.04$, which is a typical value for comets \citep[the geometric albedo of comet 67P/Churyumov-Gerasimenko is $0.05$;][]{Sierks2015}.
\par
\begin{figure*}[t!]
    \centering
    \includegraphics[angle=180,width=1.0\textwidth]{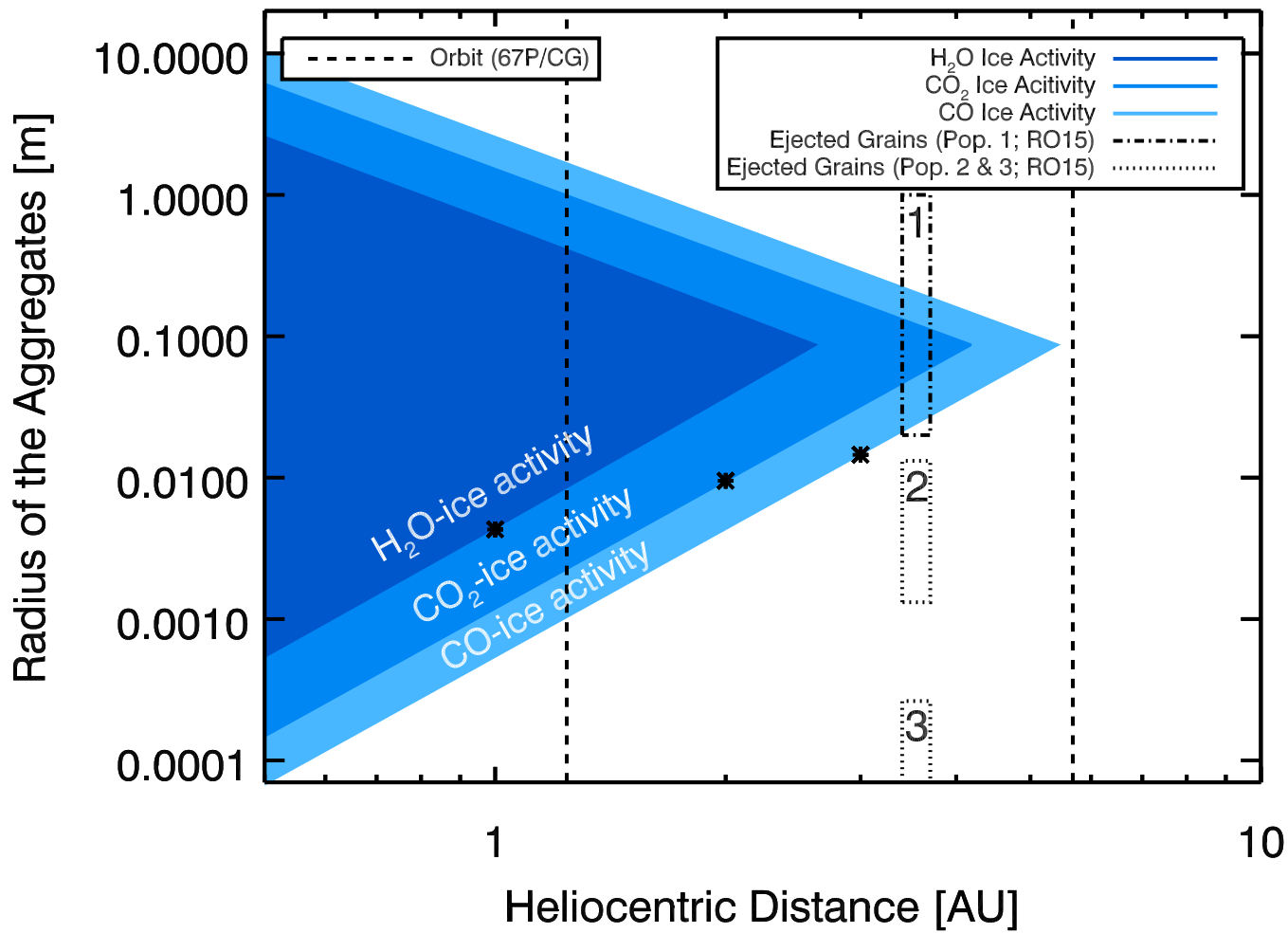}
    \caption{Radii of the dust aggregates whose adhesion can be overcome by the gas pressure (see Eqs. \ref{eq_lower} and \ref{eq_pressure}) and which are able to escape from the gravity field of the nucleus (see Eq. \ref{eq_upper}) of comet 67P/Churyumov-Gerasimenko as a function of heliocentric distance. The different blue colors denote the size range of the dust aggregates that are able to escape from the comet in case of outgassing of $\mathrm{H_2O}$ ice, $\mathrm{CO_2}$ ice, and $\mathrm{CO}$ ice. The calculations are performed for a radius of the nucleus of $2 \, \mathrm{km}$ \citep{Lowry2012}, an albedo of $0.04,$ and a density of $470 \, \mathrm{kg\, m^{-3}}$ \citep{Sierks2015}. For comparison, the three particle populations (indicated by the numbers 1 to 3; see main text for details) measured by the Grain Impact Analyzer and Dust Accumulator (GIADA) and the OSIRIS Narrow Angle Camera (NAC) are shown by the dashed-dotted box and the dotted boxes \citep{Rotundi2015}. The smallest particle size of population 3 is $28 \, \mathrm{\mu m}$ and is not shown in this figure). To derive the smallest aggregate size, we used a constant ice temperature that was determined by the exact thermophysical model at $1\,\mathrm{AU}$ for $\mathrm{H_2O}$ ice ($208.4 \, \mathrm{K}$; see caption of Fig. \ref{fig_pressures}), at $2\,\mathrm{AU}$ for $\mathrm{CO_2}$ ice ($117.2 \, \mathrm{K}$), and at $3\,\mathrm{AU}$ for $\mathrm{CO}$ ice ($33.9 \, \mathrm{K}$; these positions are indicated by the stars).}
    \label{fig_CG}
    \end{figure*}

Figure \ref{fig_CG} shows the derived dust-aggregate sizes that can be released from the cometary surface and ejected into space by the sublimation of the volatile components as a function of heliocentric distance. The blue colors visualize at which heliocentric distances the dust activity can be expected with the outgassing-activity of the different ices. For comparison, the aphelion and perihelion of comet 67P/Churyumov-Gerasimenko are denoted by the vertical dashed lines.
\par
As can be seen in Fig. \ref{fig_CG} and as was reported in \citet{Blum2015}, outgassing of $\mathrm{H_2O}$ ice can only explain the dust activity within $2.5 \, \mathrm{AU}$. At larger heliocentric distances, the outgassing rate of $\mathrm{H_2O}$ ice is not sufficient to release dust aggregates off the surface, hence, other more volatile materials must drive the dust activity at heliocentric distances larger than $2.5 \, \mathrm{AU}$.
\par
$\mathrm{CO_2}$ ice and $\mathrm{CO}$ ice have a lower latent heat than $\mathrm{H_2O}$ ice, therefore the saturation gas pressure of more volatile ices is higher than the corresponding $\mathrm{H_2O}$  ice pressure if the energy available for the sublimation is the same. Thus, at a fixed heliocentric distance, the activity of $\mathrm{CO_2}$ ice or $\mathrm{CO}$ ice can release smaller particles than the outgassing of $\mathrm{H_2O}$ ice (see the different blue areas in Fig. \ref{fig_CG} for details). Additionally, the higher gas pressure caused by the sublimation of $\mathrm{CO_2}$ ice and $\mathrm{CO}$ ice allows larger dust aggregates to escape from the gravity field of the nucleus. Thus, the size range of released particles expands in both directions.
\par
The outgassing of $\mathrm{CO_2}$ ice and $\mathrm{CO}$ ice can explain the dust activity at a heliocentric distance of up to $\sim 4\, \mathrm{AU}$ and $\sim 5.5\, \mathrm{AU}$, respectively. Thus, the gas-driven dust activity of comets is in general possible at large heliocentric distances. However, the size range of the released dust aggregates becomes smaller with increasing distance to the Sun and ends with a dust-aggregate size of $\sim 1 \, \mathrm{dm}$ close to the heliocentric distance at which the gas-driven dust activity vanishes.
\par
To explain a sustained activity at larger heliocentric distances ($> 5.5\, \mathrm{AU}$), other more volatile materials such as $\mathrm{N_2}$ ice (latent heat of sublimation\footnote{http://encyclopedia.airliquide.com/Encyclopedia.asp?GasID=5}: $\Lambda=2.2\times 10^5 \, \mathrm{J \, kg^{-1}}$) or $\mathrm{O_2}$ ice (latent heat of sublimation: $\Lambda=2.3\times 10^5 \, \mathrm{J \, kg^{-1}}$) must be present close to the surface of the nucleus (for comparison, the latent heat of sublimation of $\mathrm{CO}$ ice is $\Lambda=3.0\times 10^5 \, \mathrm{J \, kg^{-1}}$).
\par
Figure \ref{fig_CG} also shows that the size range of the dust aggregates that can be released off the nucleus widens when the comet approaches the Sun, because the pressure at the ice-dust interface increases and thereby smaller dust aggregates can be released from the surface and larger dust aggregates are capable to leave the gravitational field of the nucleus.
\par
Recent observations performed by the Rosetta spacecraft (between $3.4\, \mathrm{AU}$ and $3.7\, \mathrm{AU}$) in the inner coma of comet 67P/Churyumov-Gerasimenko revealed different populations of particles close to the nucleus \citep{Rotundi2015}. Relatively large grains (between $2\,\mathrm{cm}$ and $1\,\mathrm{m}$ in radius; see the dashed-dotted box indicated with number 1 in Fig. \ref{fig_CG}) in bound orbits have been detected by the OSIRIS Narrow Angle Camera (NAC). Figure \ref{fig_CG} shows that the outgassing of $\mathrm{CO}$ ice can explain the presence of dust aggregates with radii in excess of $~2\,\mathrm{cm}$. However, larger aggregates ($\gtrsim 3\,\mathrm{dm}$) cannot escape from the gravity field of the comet at $3.4\, \mathrm{AU}$ or $3.7\, \mathrm{AU}$. The larger aggregates can have been released closer to the Sun (during former perihelion passages). For example, the ejection of meter-sized dust aggregates is only possible close to perihelion. At perihelion of comet 67P/Churyumov-Gerasimenko, the derived size range fits the size range observed between $3.4\, \mathrm{AU}$ and $3.7\, \mathrm{AU         }$ relatively well.
\par
The second population of particles detected by the OSIRIS NAC possesses radii between $1.3\,\mathrm{mm}$ and $1.3\,\mathrm{cm}$ (see the dotted box indicated with number 2 in Fig. \ref{fig_CG}). Additionally, the third population measured by the Grain Impact Analyzer and Dust Accumulator (GIADA) consist of smaller particles with radii ranging from $28 \, \mathrm{\mu m}$ to $262 \, \mathrm{\mu m}$ (see the dotted box indicated with number 3 in Fig. \ref{fig_CG}; the lower end of this particle population is not shown in the plot). The gap between these two populations (2 and 3) is most probably caused by the detection limits of the two different instruments used to measure the particle size. This would imply that only parts of the particle size distribution in the inner coma of comet 67P/Churyumov-Gerasimenko have been investigated with these two instruments.
\par
The presented model of the gas-driven dust activity cannot explain the release of the particles found in populations 2 and 3. The release of small dust aggregates (below the blue areas shown in Fig. \ref{fig_CG}) or of micrometer-sized monomer grains is not possible because of the relatively high tensile strength required to detach these particles from the surface \citep[$\sim 1 \, \mathrm{kPa}$; see][]{Blum2014}. Thus, the particles found in populations 2 and 3 must be produced by disintegration of  previously ejected larger dust aggregates. We recall that the calculations shown in Fig. \ref{fig_CG} are only valid for dust aggregates. For $\mathrm{H_2O}$ ice aggregates or mixtures of dust and \textbf{ice}, the density of the aggregates decreases and the tensile strength between the aggregates most proabably increases, which slightly changes the result. However, if the Rosetta spacecraft were to detect icy particles, this would indicate that most probably $\mathrm{CO}$ ice or $\mathrm{CO_2}$ ice drives the dust activity (the term dust includes $\mathrm{H_2O}$ ice at this heliocentric distance) of comet 67P/Churyumov-Gerasimenko.
\par
The question remains which physical process is responsible for the production of the smaller dust-size fractions observed in comet 67P/Churyumov-Gerasimenko and in the comae and tails of other comets. Because of the estimated high internal tensile strengths of the released dust aggregates of $\gtrsim 100 \, \mathrm{Pa}$ \citep{Blum2006a}, the stress required for the disintegration or fragmentation must exceed the tensile strength. Of the three obvious processes, (1) collisions among the dust aggregates, (2) electric charging of the dust aggregates, and (3) spin-up and centrifugal mass loss of the dust aggregates, respectively, collisions seem to be too infrequent, and a typical electrostatic potential in the interplanetary space of $5 \, \mathrm{V}$ is far too low to reach the required stresses. Thus, only rotational spin-up can be considered as a candidate for mass loss of the dust aggregates on their departure from the nucleus. Assuming spherical homogeneous dust aggregates of radius $s$ and mass density $\rho_{\rm aggregate}$, spinning at a surface velocity $v$, the centrifugal stress along the equatorial region of their surface is
\begin{equation}\label{cent1}
    \sigma \, = \, \frac{m \, v^2}{s \, A} \, \mathrm{,}
\end{equation}
with $m$ and $A$ are the mass and cross section of the part of the aggregate considered to be fragmented off. Assuming the latter to be spherical with radius $s_0$ and a mass density $\rho_0$, we derive\begin{equation}\label{cent2}
    \sigma \, = \, \frac{4 \, s_0}{3\,s} \, \rho_0 \, v^2 \, \approx \, 1.3 \, \frac{s_0}{s}\, \left( \frac{v}{\mathrm{1 \, m \, s^{-1}}}\right)^2 \, \mathrm{kPa}\, \mathrm{,}
\end{equation}
with $\rho_0 =  \rho_{\rm aggregate} \approx  1000 \, \mathrm{kg \, m^{-3}}$. For a spinning velocity of $v  \approx 1 \,\mathrm{m\, s^{-1}}$ and $s_0 \approx s$, this stress approximately equals the typical internal strength of the aggregates. For much higher spinning velocities of $v \approx 100 \, \mathrm{m\, s^{-1}}$, micrometer-sized monomer grains with tensile strengths of $10^{5} \, \mathrm{Pa}$ \citep{Heim1999} can be released if $s_0 / s \gtrsim 10^{-3}$, that is, if $s \gtrsim 1 \, \mathrm{mm}$. Future studies should therefore concentrate on the gas-friction dynamics of the released dust aggregates with respect to the highest spin-up velocities. First empirical data on the rotation rate of emitted dust particles for comet 67P/Churyumov-Gerasimenko have been presented by \citet{Fulleetal2015}, who analyzed the light curves of more than 1,000 dust grains. They found that the most likely spin frequency of the particles is below 0.15 Hz, with only a few aggregates rotating at a rate higher than 1 Hz. They concluded that their observed particles had sizes in the millimeter range so that the highest rotation velocity was $v \approx 0.01 \,\mathrm{m\, s^{-1}}$. Thus, following Eq. \ref{cent2}, the highest centrifugal stress is $\sigma_{\mathrm{max}} \approx 0.1$~Pa, much too low to explain the formation of small dust aggregates by centrifugal splitting.

\section{Summary and discussion}\label{Summary}
The main findings of this work can be summarized as follows
\begin{itemize}
\item We developed a simplified analytical model to derive the size of the dust aggregates that can be released from the cometary surface into space by the sublimation of volatile materials covered by a nonvolatile surface layer. This model takes the cohesion of the material into account, which yields, together with the gas pressure at the ice-dust interface, the smallest size of the dust aggregates. Additionally, the gravitational force in combination with the gas friction force determines the largest size of dust aggregates that are able to escape from the gravity field of the comet. Our model is a one-species model, which means that all incoming energy is consumed by the sublimation of only
one volatile component (in reality, comets are multicomponent mixtures with energy partitioning among the different volatile species). Furthermore, this model is based on the assumption that comets have formed by gravitational collapse of dust agglomerates in the early solar system \citep{Johansen2007,Skorov2012,Blum2014}, which implies that the surface of the nucleus consists of dust aggregates (the aggregates themselves are composed of micrometer-sized grains). We showed that the simplified analytical model agrees
well with numerical calculations using a sophisticated thermophysical model (see Fig. \ref{fig_pressures} in Sect. \ref{Modeling}).
\item The simplified model was then applied to comet 67P/Churyumov-Gerasimenko, taking into account three different volatile materials: $\mathrm{H_2O}$ ice, $\mathrm{CO_2}$ ice, and $\mathrm{CO}$ ice (see Fig. \ref{fig_CG} in Sect. \ref{Application}). $\mathrm{H_2O}$ ice can explain the gas-driven dust activity of comet 67P/Churyumov-Gerasimenko within a heliocentric distance of $2.5 \, \mathrm{AU}$. For $\mathrm{CO_2}$ ice and $\mathrm{CO}$ ice, the gas-driven dust activity is possible up to $\sim 4\, \mathrm{AU}$ and $\sim 5.5\, \mathrm{AU}$, respectively. Thus, outgassing of $\mathrm{H_2O}$ ice cannot drive the observed dust activity between $3.4\, \mathrm{AU}$ and $3.7\, \mathrm{AU}$ \citep[position of comet 67P/Churyumov-Gerasimenko during the first dust size measurements in the inner coma;][]{Rotundi2015}.
\item The size range of the dust aggregates able to escape from the nucleus into space widens when the comet approaches the Sun and narrows with increasing heliocentric distance (see Fig. \ref{fig_CG} in Sect. \ref{Application}). Independent of the volatile material, the size range always ends at larger heliocentric distances with an aggregate radius of $\sim 1 \, \mathrm{dm}$, which is also the observed size of cometary trails \citep[see, e.g.,][]{Reachetal2000, Harmonetal2004, Nolanetal2006, Ishiguro2008, Trigoetal2010}. The size range of the dust aggregates widens when the comet approaches to the Sun because the tensile strength of the dust aggregates decreases with increasing aggregate size. Thus, with higher pressure available at the ice-dust interface closer to the Sun, smaller dust aggregates can be released from the surface. Furthermore, the pressure increase enables larger dust aggregates to escape against the gravity of the nucleus into space. This can explain the presence of the aggregates larger than $\sim 3 \, \mathrm{dm}$ if these aggregates have been released during former perihelion passages. Recently, \citet{Sojaetal2015} found that the trail of comet 67P/Churyumov-Gerasimenko is dominated by millimeter-sized particles. Our model can explain the release of these particles, but only with the activity of CO ice close to perihelion.
\item The model can explain the presence of bound grains \citep[with radii between $2\,\mathrm{cm}$ and $1\,\mathrm{m}$ ;][]{Rotundi2015} close to the nucleus of comet 67P/Churyumov-Gerasimenko (dashed-dotted box in Fig. \ref{fig_CG}). However, the smaller particles detected by GIADA and OSIRIS NAC onboard the Rosetta spacecraft at heliocentric distances between $3.4\, \mathrm{AU}$ and $3.7\, \mathrm{AU}$ \citep[dotted boxes in Fig. \ref{fig_CG};][]{Rotundi2015} and between $2.4\, \mathrm{AU}$ and $3.4\, \mathrm{AU}$ \citep{DellaCorteetal2015} cannot be released by the gas pressure at these heliocentric distances. These particles must be the product of a disintegration of larger dust aggregates in the coma. Furthermore, our model roughly agrees with the findings by \citet{Fulleetal2015}, showing that particles with radii of a few millimeters are a good model for the OSIRIS observations of free-floating particles in the coma.
\item The release of micrometer-sized monomer grains is not possible because of the relatively high tensile strength required to detach the grains from the surface \citep[$\sim 1 \, \mathrm{kPa}$; see][]{Blum2014}. This high gas pressure cannot be reached at the ice-dust interface under normal conditions.
\end{itemize}
The results we discussed can also be used to model the gas-driven dust activity of other comets. The application of our simplified model to another comet would shift the largest dust-aggregate size (upper limitation of the blue areas in Fig. \ref{fig_CG}) and thereby the heliocentric distance at which the dust activity starts.

\begin{acknowledgements}
We thank M. Mumma for an interesting discussion about volatile species and the anonymous referee for helpful comments.
\end{acknowledgements}

\bibliographystyle{aa}

\bibliography{bib}

\end{document}